\documentclass[conference]{IEEEtran}

\usepackage{comment,cite}
\usepackage{amsmath, amssymb}
\usepackage{bbm}
\usepackage{balance}

\usepackage{microtype}
\usepackage{graphicx}
\usepackage{subfigure}
\usepackage{booktabs} 

\DeclareMathOperator*{\argmax}{argmax}

\usepackage{mathtools}
\DeclarePairedDelimiter\ceil{\lceil}{\rceil}

\newcommand{\myapprox}{{\raise.17ex\hbox{$\scriptstyle\sim$}}}

\usepackage{xcolor}
\usepackage{braket, amsthm, amsfonts}
\usepackage[ruled,longend]{algorithm2e}
\usepackage{hyperref}

\usepackage{multirow}


\newlength\myindent
\newcommand{\E}{\mathbb{E}}
\usepackage[font=small,skip=0pt]{caption}

\makeatletter
\g@addto@macro\normalsize{%
  \setlength\abovedisplayskip{5pt}
  \setlength\belowdisplayskip{5pt}
  \setlength\abovedisplayshortskip{5pt}
  \setlength\belowdisplayshortskip{5pt}
}
\makeatother

\newtheorem{approximation}{Approximation}
\newtheorem{claim}{Claim}

\makeatother

\begin{document}

\title{Optimizing Redundancy Levels in Master-Worker Compute Clusters for Straggler Mitigation}

\author{%
  \IEEEauthorblockN{Mehmet Fatih Akta\c{s} and Emina Soljanin} \\
  \IEEEauthorblockA{Department of Electrical and Computer Engineering, Rutgers University \\
  Email: \{mehmet.aktas, emina.soljanin\}@rutgers.edu}
}
\maketitle

\begin{abstract}
Runtime variability in computing systems causes some tasks to straggle and take much longer than expected to complete. These straggler tasks are known to significantly slowdown distributed computation.
Job execution with speculative execution of redundant tasks has been the most widely deployed technique for mitigating the impact of stragglers, and many recent theoretical papers have studied the advantages and disadvantages of using redundancy under various system and service models.
However, no clear guidelines could yet be found on when, for which jobs, and how much redundancy should be employed in Master-Worker compute clusters, which is the most widely adopted architecture in modern compute systems.
We are concerned with finding a strategy for scheduling jobs with redundancy that works well in practice.
This is a complex optimization problem, which we address in stages.
We first use Reinforcement Learning (RL) techniques to learn good scheduling principles from realistic experience.
Building on these principles, we derive a simple scheduling policy and present an approximate analysis of its performance.
Specifically, we derive expressions to decide when and which jobs should be scheduled with how much redundancy.
We show that policy that we devise in this way performs as good as the more complex policies that are derived by RL.
Finally, we extend our approximate analysis to the case when system employs the other widely deployed remedy for stragglers, which is relaunching straggler tasks after waiting some time.
We show that scheduling with redundancy significantly outperforms straggler relaunch policy when the offered load on the system is low or moderate, and performs slightly worse when the offered load is very high.
\end{abstract}


\section{Introduction}
Large-scale compute jobs, such as those executing complex machine learning (ML) algorithms, are split into multiple tasks that are executed in parallel over distributed resources.
Tasks running in modern compute clusters have been shown to exhibit significant variability in their execution times \cite{AttackOfClones:AnanthanarayananGS13, MapReduce:DeanG08, ImprovingMapReducePerformance:ZahariaKJ08, ReiningOutliersInMapReduce:AnanthanarayananKG10, TailAtScale:DeanB13, AvoidingLatencyTraps:XuBN13, StragglerRootCause:GarraghanOY16}.
Runtime variability randomly causes some tasks run slow, which is commonly referred to as \emph{straggling}.
A distributed job finishes only when its slowest task completes, and as the number of tasks within a job increases, so does the chance that at least one of them will be  a straggler. Because of that, stragglers have become a great concern for today's large-scale compute workloads \cite{TailAtScale:DeanB13}.

\emph{Redundancy} has long been used in production systems as a tool to attain predictable performance at the presence of runtime variability \cite{AmazonDynamo:DecandiaHJ07, MapReduce:DeanG08}.
The idea is to speculatively launch multiple copies for the same task and wait only for the fastest one to complete, hence avoid stragglers.
Task replication has been shown to effectively mitigate stragglers both in practice \cite{TailAtScale:DeanB13, LowLatencyviaRed:VulimiriGM13, SGDWithReplicas:ChenPM16} and theory \cite{Codes&Qs:JoshiLS12, Codes&Qs:HuangPZ12, CodesQs:KadheSS15_Allerton, Redd:Gardner17}.
Replica tasks bring additional load on the system, thus the guidance for employing them is usually conservative; replicas are launched only for ``short'' tasks \cite{AttackOfClones:AnanthanarayananGS13}, or only for tasks that seem to straggle \cite{ImprovingMapReducePerformance:ZahariaKJ08}, or proposed to be issued only to idle servers \cite{MASCOTS:GardnerHS16}.
\emph{Erasure coding} implements a more general form of redundancy than replication and has been shown to mitigate stragglers by introducing smaller redundant load on the system \cite{MAMA:AktasPS17, IFIP:AktasPS17}.
Coding techniques have been applied for straggler tolerance in common distributed linear computation \cite{ShortDot:DuttaCG16, CodedConvolution:DuttaCG17, CodedMatrixMultiplication:YuMA18} or iterative optimization algorithms \cite{MachineLearningWithCodes:LeeLP17, CodedGradientDescent:LiKA17, GradientCoding:RavivTT17, GradientCoding:TandonLD17, StragglerMitigationWithDataEncoding:KarakusSD17, CodedGradientDescent:HalbawiAS18} that empower large scale ML.

Despite the plethora of papers devising new redundancy techniques, no clear guidelines could yet be found on how to schedule jobs with redundancy.
Practitioners currently resort to heuristics such as scheduling only ``short'' jobs with redundancy. However, even then important questions have yet to be addressed.
Jobs arriving to a compute cluster consist of varying number of tasks, request varying amount of resource and have random service times. How should we quantify the \emph{total demand} of a job? Which jobs are short enough to be scheduled with redundancy?
When employed excessively, redundancy may aggravate the job slowdowns, or even cause early instability \cite{AttackOfClones:AnanthanarayananGS13, DecouplingSlowdownAndJobSize:Gardner17}.
At what level of system load does redundancy start to hurt performance? If a job is going to be scheduled with redundancy, how much redundancy should be embedded into its execution?

Answers for the questions posed above depend on many factors, most importantly, on the system architecture, job sizes and requirements, offered load on the cluster, and characteristics of runtime variability.
Ideally, an analytic understanding of the relationship between the important decision making parameters and the system performance would reveal great deal of insight, and serve as an excellent tool for designing a good policy for scheduling with redundancy.
Performance of systems with redundancy has been studied analytically under various system models and assumptions \cite{QueuesWithRed:JoshiSW15, RedForCloud:Joshi17, Redd:Gardner17, DeltaProbingPoliciesForRed:RaaijmakersBB18}.
Major challenge in this pursuit is that \emph{performance analysis of systems with redundancy proved to be intractable}, even under simplified settings that assume single-task job arrivals and servers which can serve one task at a time \cite{QueuesWithRed:JoshiSW15, RedForCloud:Joshi17, Redd:Gardner17, DecouplingSlowdownAndJobSize:Gardner17, DeltaProbingPoliciesForRed:RaaijmakersBB18}.
The only known exact analysis for systems with redundancy has been presented in \cite{Redd:Gardner17} for a queueing system with a randomized scheduling of task replicas under the assumption that jobs consist of only one task, and their service times are exponential and independent across servers.
Same authors introduced in \cite{MASCOTS:GardnerHS16} a better model that decouples the runtime variability from the inherent task sizes, and show that the updated model supports the experimental fact that excessive redundancy hurts performance. However, as noted by the authors, giving up on the independent exponential service time model renders the exact analysis of systems with redundancy to be simply formidable.

This paper considers the problem of scheduling with redundancy in Master-Worker architecture. Despite its wide adoption in modern compute systems (e.g., Apache Hadoop~\cite{HadoopYarn:VavilapalliMD13}, Kubernetes~\cite{BorgOmegaKubernetes:Burns16}, Mesos~\cite{Mesos:HindmanKZ2011}), scheduling for straggler mitigation, to the best of our knowledge, has not been theoretically studied for this architecture.
Research on scheduling with redundancy has so far considered simplified models such as compute jobs consisting of only a single task and compute nodes being able to serve only a single task at a time.
We use as few simplifying assumptions as possible in our system model to find the scheduling policy that performs well in practice.
Specifically in our model, arriving jobs consist of random number of tasks, request random amount of resource capacity, take random amount of time to complete execution. We adopt the straggler model developed in \cite{MASCOTS:GardnerHS16} according to which runtime variability expands task service times by a random multiplicative factor.

We address the problem of scheduling compute jobs with redundancy with a combination of Reinforcement Learning (RL) and mathematical modeling.
Scheduling is a control problem, and RL techniques have recently been applied and generated insight into scheduler design on various problems, e.g., \cite{ResManWithDeepRL:Mao16, KnowledgeDefinedNetworking:Mestres17, MLForRouting:Valadarsky17, TaskPlacementOnWANWithDRL:Wang17, DevicePlacementWithDRL:Mirhoseini17}.
Inspired by these successful RL applications, we firstly use RL techniques to learn from realistic experience the principles for effective scheduling of redundancy.
Specifically with Deep Q-learning, we learn that \emph{right amount} of redundancy shall be introduced in executing \emph{small enough} jobs, and only when the cluster operates under \emph{low enough} load.

RL techniques are useful to derive good scheduling principles but they suffer from the well known shortcomings such as: 1) require many training hours to converge, 2) may get trapped in local optima, 3) results may not generalize when the learning environment slightly changes \cite{RLDoesNotWorkYet:Irpan18}.
Building on the principles that are learned by Deep-RL, we propose a simpler policy $\mathrm{Redundant\textnormal{-}small}$ that schedules jobs with redundancy only if their \emph{total demand} is below $d$.
With mathematical modeling, we derive an approximation for the system and show that it is able to predict the simulated average system response time fairly accurately.
Most importantly, approximate expressions that we derive allow tuning $d$ in order to maximize the performance of $\mathrm{Redundant\textnormal{-}small}$.
We conclude that $\mathrm{Redundant\textnormal{-}small}$ derived using our approximation performs as good as the more complex policies derived by Deep-RL.

We finally consider \emph{straggler relaunch}, which is another widely deployed remedy for stragglers \cite{ImprovingMapReducePerformance:ZahariaKJ08}.
In particular, we study the performance of Master-Worker cluster under $\mathrm{Straggler\textnormal{-}relaunch}$ policy that sets a timer for the tasks within each job at the time of scheduling, then cancels and relaunches a task once its timer expires.
We extend the approximate analysis that we derive for the system with redundancy to the system with straggler relaunch. Our analysis allows optimizing the amount of waiting time before performing relaunch for the tasks served in the system.
Comparing the performance of optimized $\mathrm{Redundant\textnormal{-}small}$ and optimized $\mathrm{Straggler\textnormal{-}relaunch}$ policies, we find that scheduling with redundancy significantly outperforms straggler relaunch when offered load on the system is low or moderate, and performs worse when the offered load is very high ($\gtrsim 0.85$).

\begin{figure}[t]
  \centering
  \includegraphics[width=0.38\textwidth, keepaspectratio=true]{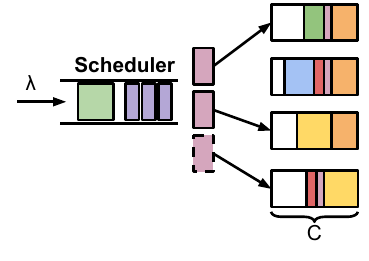}
  \caption{System model for scheduling with redundancy. A job of two tasks (solid) gets scheduled with a redundant task (dashed) such that any two of the three tasks is sufficient for its completion.}
  \label{fig:fig_kubernetes}
\end{figure}

\section{System Model}
\label{sec:sec_sys_model}
\noindent
\textbf{System architecture and job arrivals:}
We consider a Master-Worker compute cluster architecture as implemented in Kubernetes; a cluster management framework widely used in production cloud systems \cite{BorgOmegaKubernetes:Burns16}. Note that Master-Worker architecture is widely deployed not only in cloud but also in modern high performance computing systems, where it is often referred to as First-come First-served batch scheduling architecture \cite{TimeSharingInHPC:HofmeyrIC16}.
A cluster consists of a single scheduler (master) managing $N$ nodes (slaves), each with capacity $C$.
Jobs arrive as a Poisson process of rate $\lambda$, each consisting of a random number ($k$) of tasks. Tasks within the same job request equal ($r$) amount of capacity and have the same minimum service time ($b$). We assume $k$, $r$ and $b$ are independently sampled from random variables $K$, $R$ and $B$ for each job.
Job service times in practice exhibit heavy tail, in particular, commonly distributed as Pareto~\cite{HeavyTailedJobs:Leland86, HeavyTailedJobs:Harchol97, GoogleClusterDataAnalysis:ChenGG10, GoogleTraceAnalysis:ReissTG12, AnAnalysisOnAlibabaClusterTrace:LuYX17}.
Thus, we model the minimum service time $B$ as a $\mathrm{Pareto}$ random variable that is characterized by its minimum value $b_{\min}$ and tail index $\beta$ as
\[ \Pr\{B > b\} = \left(b_{\min}/b\right)^{\beta} \quad \mathrm{for}~~ b > b_{\min}. \]
Note that it is more appropriate to model service times using an upper truncated Pareto distribution. We choose to not use the truncated version since we get the same results under either model as long as the ratio between the maximum and minimum values of the truncated Pareto is sufficiently large, which is known to be the case in real compute jobs \cite{HeavyTailedJobs:Harchol97, GoogleTraceAnalysis:ReissTG12, AttackOfClones:AnanthanarayananGS13}.
As discussed later in Sec.~\ref{sec:sec_learning}, distribution of $R$ turned out to be not significant to find a good policy for scheduling with redundancy. For this reason and to keep the discussion simpler throughout the paper, we set $R=1$. Note that this is not a limiting assumption and the study that we present easily extends to the case with random $R$.

Number of tasks in real compute jobs has also been shown to exhibit heavy tail \cite{GoogleClusterDataAnalysis:ChenGG10, GoogleTraceAnalysis:ReissTG12}.
One canonical discrete heavy tailed distribution is Zipf, which we adopt here by modeling the number of tasks $K$ as a $\mathrm{Zipf}$ random variable with an exponent of $1$ and a maximum value of $k_{\max}$
\[ \Pr\{K = k\} = \frac{1/k}{\sum_{i=1}^{k_{\max}} 1/i} \quad \mathrm{for}~~ k = 1, \dots, k_{\max}. \]

\vspace{0.5em}
\noindent
\textbf{Runtime variability:}
We model runtime variability with a random variable $S$ that is identically and independently distributed (i.i.d.) across different nodes and tasks.
Once a task with service time $b$ starts execution, it samples a straggling factor $s > 1$ from $S$ and takes $s \times b$ of time to complete. This model is introduced in \cite{DecouplingSlowdownAndJobSize:Gardner17} and shown to support the experimental evidence.
To capture the significant variability that is observed in practice, we model $S$ as a $\mathrm{Pareto}$ with a minimum value of $1$ and tail index $\alpha$.

\vspace{0.5em}
\noindent
\textbf{Scheduling:}
We adopt the scheduling dynamics implemented in Kubernetes (see Fig.~\ref{fig:fig_kubernetes}).
Scheduler is continuously updated with the resource availability at each cluster node. Number of tasks and their requested resource capacity are known for each arriving job. Additionally, we assume task service times are also known.
Jobs wait in a first-in first-out queue to get scheduled according to a work-conserving policy; job at the head of the queue gets dispatched as soon as enough resources become available in the cluster to fit all its tasks (both initial and redundant).
Scheduler distributes the offered load evenly across the nodes; tasks of a job are assigned to the least loaded nodes among all with sufficient available capacity.

Scheduler decides how many redundant tasks to embed into each arriving job.
Job execution with coded redundancy has been shown to be more effective than replication for straggler mitigation \cite{MAMA:AktasPS17, IFIP:AktasPS17}.
Therefore, we here focus on scheduling \emph{coded redundancy}, but the presented study can be directly extended for replicated redundancy.
With coding, a job of $k$ tasks is expanded into a job of $n$ tasks by embedding $n-k$ \emph{parity} tasks into its execution. Parity tasks are constructed by encoding the initial $k$ tasks, either by adding redundancy in the computational procedure that the tasks collaboratively implement (e.g., \cite{ShortDot:DuttaCG16}) or inserting redundancy in the data that the tasks consume (e.g., \cite{StragglerMitigationWithDataEncoding:KarakusSD17}).
We consider the most commonly used encoding model, \emph{MDS} codes, under which executing any $k$ of the $n$ tasks is sufficient to recover the desired job outcome. As soon as $k$ fastest tasks of the job complete service, the remaining outstanding $n-k$ tasks will be removed from service, which is assumed to cause no extra delay.

\vspace{0.5em}
\noindent
\textbf{System configuration:}
We built a cluster simulator using SimPy \cite{SimPy:Matloff08} to implement the system model described above\footnote{We made the cluster simulator and our implementation of the Deep-Q learning Algorithm~\ref{alg:alg_q_learning} available on \url{github.com/mfatihaktas/deep-scheduler}.}.
Results presented in the plots presented throughout the paper are generated by setting the system parameters as $N=20$, $C=10$, $k_{\max}=10$, $b_{\min}=10$, $\beta=3$, $\alpha=3$, and varying the arrival rate $\lambda$ to change the offered load on the cluster.
Reported simulation results are generated by sampling from 30 different runs, where each simulation run is executed until the first 100,000 job arrivals finish execution.

\vspace{0.5em}
\noindent
\textbf{Notation and Tools for Analysis:}
We here give an overview of the notation and special functions that appear throughout the paper.
For their detailed definitions and interesting properties, we refer the reader to \cite{NIST:DLMF}.
We denote the expectation with respect to a random variable $X$ as $\E_X$.
$X_{n:i}$ denotes the $i$th order statistic of $n$ i.i.d. samples drawn from a random variable $X$.
Incomplete Beta function $B(q;m,n)$ is defined for $q \in [0,1]$, $m, n \in R^+$ as $\int_0^q u^{m-1}(1-u)^{n-1} du$, Beta function $B(m,n)$ as $B(1;m,n)$ and its regularized form $I(q;m,n)$ as $B(q;m,n)/B(m,n)$.
Gamma function $\Gamma(x)$ is defined as $\int_0^{\infty} u^{x-1}e^{-u}du$ for $x \in R$ or as $(x-1)!$ for $x \in Z^+$.

\section{Learning how to schedule with redundancy}
\label{sec:sec_learning}
\noindent
\textbf{RL formulation for scheduling with redundancy:}
We use model-free RL that considers an \emph{agent} interacting with a previously unknown \emph{environment}. At each time step, agent observes a \emph{state}, executes an \emph{action}, and collects a \emph{reward} for each of its executed state-action pairs. Actions are generated according to a \emph{policy} and RL is concerned with finding a good policy to achieve a high \emph{cumulative reward}.

In our problem, environment is a compute cluster and agent is the scheduler. Scheduler interacts with the system by embedding redundancy to arriving jobs and assigning their tasks on to the cluster nodes (see Sec.~\ref{sec:sec_sys_model} for how).
We set scheduler's goal as to \emph{minimize job slowdowns}.
This is because job slowdown relates the total time a job spends in the system to job's minimum service time, hence has long been suggested to be a better performance evaluation metric than others \cite{FirstMentionOfJobSlowdown:Hansen71, PerfMetricsForParallelJobScheduling:Feitelson01}.
Precisely, the slowdown experienced by a job is defined as the total time it actually spends in the system divided by its minimum service time.
Note that system performance analysis in terms of job slowdowns with mathematical modeling is known to be often formidable, even adopting very simplified models \cite{HeavyTailedJobs:Harchol97}. 
RL, however, is oblivious to the performance metric that is used while searching for a good policy.

While scheduling a job, we feed two state inputs to the scheduler: (i) average load on the cluster nodes that job's tasks are assigned to and (ii) job's \emph{demand}, which is defined as $k \times r \times b$ where $r$ is the requested resource capacity and $b$ is the minimum service time for each of the $k$ tasks in the job.
Note that scheduler can in reality access more information in the cluster, such as the job queue length, the load at each cluster node, or $k$, $r$ and $b$ separately for each job (recall the Kubernetes model in Sec.~\ref{sec:sec_sys_model}).
In our experiments, expanding the RL state with such detail did not improve job slowdowns and significantly slowed down the policy learning process.

Scheduler decides (acts) on the number of redundant tasks to embed into each job at the time of scheduling.
Reward is the signal returned by the system to each state-action executed by the scheduler and should be properly crafted to guide the policy search towards minimizing job slowdowns.
We use the negative of the slowdown a job experiences as the reward for its scheduling action.

\vspace{0.5em}
\noindent
\textbf{RL implementation:}
Foundations of RL are laid by the framework of Markov Decision Processes (MDPs) \cite{RL:Sutton98}.
Given an MDP with a fixed policy $\pi$, value $V_{\pi}(s)$ of being at state $s$ is defined as the cumulative reward expected by following $\pi$ from that point on. Similarly, value $Q_{\pi}(s, a)$ of taking action $a$ at state $s$ is defined as $r + V_{\pi}(s^+)$, where $r$ is the reward collected for $s$-$a$ and $s^+$ is the next state.
For every MDP, there is at least one optimal policy $\pi^*$ such that $V_{\pi^*}(s) > V_{\pi}(s)$ at every state $s$.
Bellman optimality equation says that greedily following the action that maximizes the \emph{optimal} Q-value at each state is an optimal policy. Thus, it is sufficient to find the optimal Q-value function $Q^*(s, a)$ to find an optimal policy.

There are two model-free RL techniques: Monte Carlo and and Temporal-Difference (TD) learning. We use the latter since it is known to be more data efficient on Markovian environments such as our system.
TD methods learn a quantity of interest (e.g., Q-values) by \emph{bootstrapping} each newly collected sample on its previous estimates, that is, updating the estimates to a weighted sum of the previous estimates and the newly collected one-step rewards and then improving the policy accordingly. 
In particular, \emph{Q-learning} is implemented by iteratively updating the Q-value estimates with the newly observed state-action-reward-next state ($s_t, a_t, r_t, s_{t+1}$) tuple at each time step $t$ as
\begin{equation}
\begin{split}
  \hat{Q}(s_t,& a_t) = \hat{Q}(s_t, a_t) \\
  &+ \alpha\left(r_t + \gamma \max_a \hat{Q}(s_{t+1}, a) - \hat{Q}(s_t, a_t)\right).
\end{split}
\label{eq:eq_qlearning_update}
\end{equation}
for $\gamma < 1$ and $\alpha > 0$. Q-learning is known to provably converge to $Q^*(s, a)$ as $t \to \infty$.

Two factors contribute to the overall slowdown experienced by a job: i) waiting time in the queue, ii) slowdown due to runtime variability. Scheduling a job with redundancy mitigates the runtime variability, but redundant tasks occupy extra resources, which is likely to increase the waiting times for the subsequent jobs.
Therefore, in order to quantify the performance (i.e., estimate Q-values) for a given scheduling policy, one needs to collect experience for a sufficiently long sequence of jobs. Collected sequence of jobs should be continuous in the scheduling order (i.e., should not skip jobs) since the impact of a scheduling action is highest on the immediately subsequent job and decays on the jobs that are further away.
In addition, shorter jobs can finish earlier and fill up the experience sequence, hence skipping jobs might cause discriminating against the longer jobs and result in not collecting enough experience for those.
Thus, we divide the learning process into \emph{episodes}, and within each we collect experience for a fixed number of subsequent jobs, then execute the Q-learning update.

We use Deep Q-learning, that is, $Q(s, a)$ is approximated with a vanilla three-layer neural \emph{Q-network}.
Deep Q-learning is known to suffer from \emph{correlations} within the training data and \emph{non-stationary} target Q-values. To stabilize the learning process, we use \emph{experience replay} and a separate \emph{Target-network} to read the Q-value estimates.
Most importantly, in order to efficiently converge to an optimal policy, a balance should be implemented between \emph{exploration} (gaining information about the environment) and \emph{exploitation} (making more rewarding state-action's more likely).
To implement this balance, we use \emph{off-policy control}, that is, learning optimal Q-values while choosing actions according to an exploratory policy. We implement the exploratory policy using \emph{Upper Confidence Bound} (UCB) algorithm (as in \cite{ComparingExplorationTechniquesInQlearning:Tijsma16}).
We discretize the state space in order to count the number of state-action visits required by UCB.
These three techniques are fairly standard and we do not discuss them in detail but display their role in Algorithm~\ref{alg:alg_q_learning}.

A pseudo-code for our Q-learning implementation is given in Algorithm~\ref{alg:alg_q_learning}, and the important steps executed within each learning episode are summarized as follows.
Scheduling decisions are made by the UCB algorithm reading Q-value estimates per state-action from the Q-network.
Within each learning episode, experience (state-action-reward-next state) gets collected for $M$ subsequently scheduled jobs and pushed into experience replay buffer. At the end of each episode, a $B$-size batch of experience is sampled uniformly at random from the replay buffer and Q-network is trained by bootstrapping on the Q-value estimates read from Target-network.
This step is repeated multiple times to learn more efficiently from the experience available in the replay buffer.
Q-network is periodically copied into Target-network.

\SetKwComment{tcp}{\small $\triangleright$ }{}%
\SetCommentSty{small}


\begin{algorithm}[!ht]
\caption{Q-learning pseudo-code: Learning how to schedule jobs with redundancy from experience.}
  Initialize $\gamma$, $\alpha$, $B$, $M$, and
  
  $~~$$\hat{Q}(s, a)$, $\hat{T}(s, a)$ \tcp*{Q-network and Target-network}
  $~~$$\mathrm{exp\_sequence}$ \tcp*{List of job experience tuples}
  $~~$$\mathrm{exp\_buffer}$ \tcp*{Experience replay queue (FIFO)}
  $~~$$N(s, a)$ \tcp*{Number of times $(s, a)$ is visited (for UCB)}
  $j = 1$ \tcp*{Id for the first job in the current episode}
  
  $i = 0$ \tcp*{Run the following two loops in parallel.}
  
  \While{true}
  {
    $i \gets i + 1$
    
    Retrieve the first from job queue and assign it with id $i$.
    
    Observe state $s_i$ and take action $a_i$ on job-$i$; \\
    $a_i = \argmax_a\; \hat{Q}(s_i, a) + \sqrt{2 \frac{\log\left(\sum_{a^\prime} N(s_i, a^\prime)\right)}{N(s_i, a)}}.$
    
    
    Store $s_i$, $a_i$ in $\mathrm{exp\_sequence}$.
    
    Discretize $s_i$, $a_i$ and increment $N(s_i, a_i)$.
  }
  \While{true}
  {
    Listen for a job completion, let its id be $i$.
    
    Store reward $r_i$\;($-\mathrm{slowdown}_i$) in $\mathrm{exp\_sequence}$.
    
    \If{all jobs with an id in $[j, j+M-1]$ are finished}
    {
      Push collected $(s_i, a_i, r_i, s_i^+)$'s in $\mathrm{exp\_buffer}$. \\
        (where next state $s_i^+=s_{i+1}$)
      
      Sample a batch of $B$ tuples from $\mathrm{exp\_buffer}$.
      
      \tcc{Repeat the following two loops several times.}
      \For{each $(s_i, a_i, r_i, s_i^+)$ in batch}
      {
        $T_i = r_i + \gamma\max_a \hat{T}(s_i^+, a)$
      }
      
      \For{each $(s_i, a_i, r_i, s_i^+)$ in batch}
      {
        $\hat{Q}(s_i, a_i) \gets \hat{Q}(s_i, a_i)
        + \alpha\left(T_i - \hat{Q}(s_i, a_i) \right)$
      }
      $j = \text{last scheduled job's id} + 1$
    }
    Periodically update $\hat{Q}$'s parameters with those of $\hat{T}$.
  }
\label{alg:alg_q_learning}
\end{algorithm}

\begin{figure}[t]
  \centering
  \begin{subfigure}
    \centering
    \includegraphics[width=.23\textwidth, keepaspectratio=true]{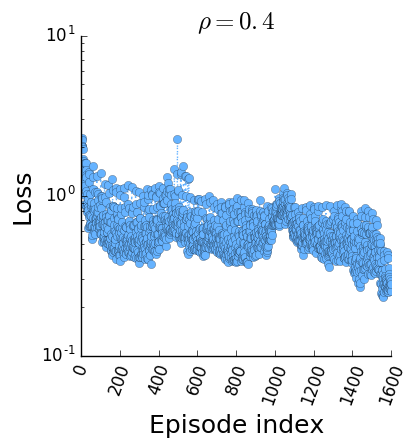}
  \end{subfigure}
  \begin{subfigure}
    \centering
    \includegraphics[width=.23\textwidth, keepaspectratio=true]{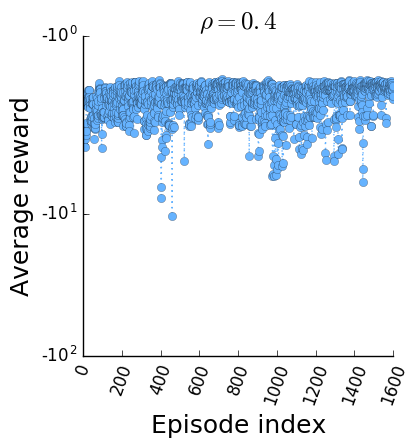}
  \end{subfigure}
  \begin{subfigure}
    \centering
    \includegraphics[width=.23\textwidth, keepaspectratio=true]{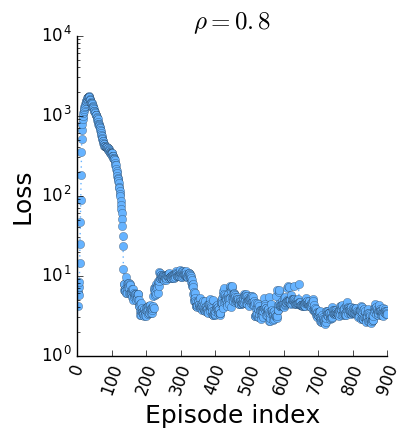}
  \end{subfigure}
  \begin{subfigure}
    \centering
    \includegraphics[width=.23\textwidth, keepaspectratio=true]{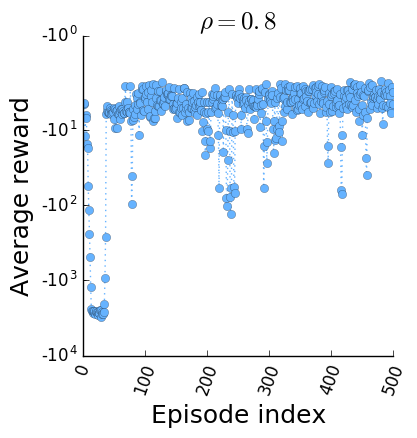}
  \end{subfigure}
  \caption{Accuracy of Q-network (Huber loss, Left) and average collected reward (Right) over the learning episodes.
  Each row of curves is generated with a single run of Algorithm~\ref{alg:alg_q_learning}.}
  \label{fig:plot_Qlearning_convergence}
\end{figure}

\vspace{0.5em}
\noindent
\textbf{RL evaluation:}
We evaluate Algorithm~\ref{alg:alg_q_learning} by running it with our cluster simulator (explained in Sec.~\ref{sec:sec_sys_model}) under different values of offered load $\rho$ on the cluster.
We here set the maximum number of redundant tasks that scheduler can embed into jobs to three.
For each different $\rho$, we run the algorithm until it converges and settles down on a policy.

\begin{figure}[thb]
  \centering
  \begin{subfigure}
    \centering
    \includegraphics[width=.23\textwidth, keepaspectratio=true]{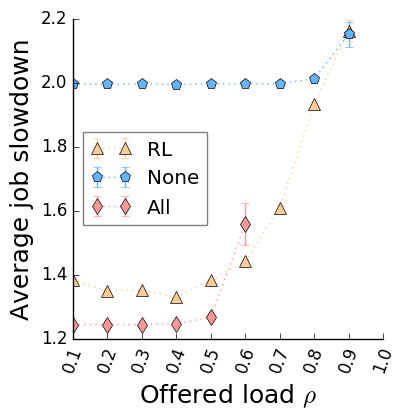}
  \end{subfigure}
  \begin{subfigure}
    \centering
    \includegraphics[width=.23\textwidth, keepaspectratio=true]{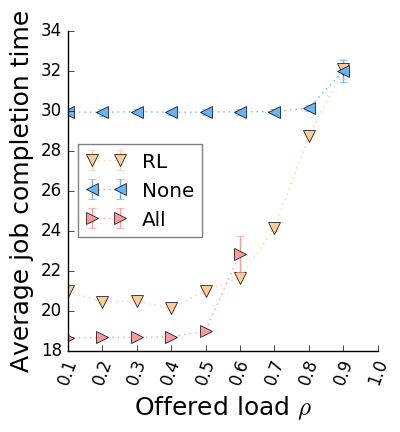}
  \end{subfigure}
  \caption{Average job slowdown and job completion times for $\mathrm{Redundant\textnormal{-}small}$ (RL), $\mathrm{Redundant\textnormal{-}all}$ (All) and $\mathrm{Redundant\textnormal{-}none}$ (None) under varying offered load.
  $\mathrm{Redundant\textnormal{-}all}$ destabilizes the system beyond $\rho=0.6$, which are excluded from the plots.}
  \label{fig:plot_Esl_ET_wlearning}
\end{figure}

\begin{figure}[t]
  \centering
  \begin{subfigure}
    \centering
    \includegraphics[width=.23\textwidth, keepaspectratio=true]{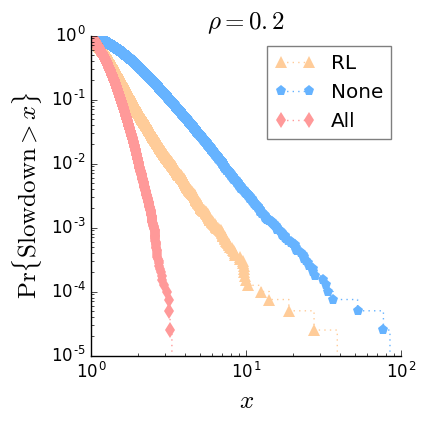}
  \end{subfigure}
  \begin{subfigure}
    \centering
    \includegraphics[width=.23\textwidth, keepaspectratio=true]{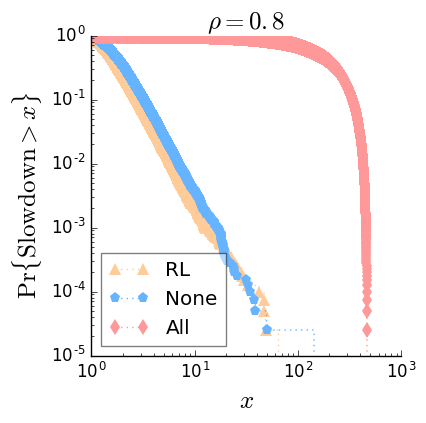}
  \end{subfigure}
  \caption{Tail distribution of job slowdowns. Each curve is sampled from a single simulation run.}
  \label{fig:plot_slowdown_tail_wlearning}
\end{figure}

\begin{figure*}[t]
  \centering
  \begin{subfigure}
    \centering
    \includegraphics[width=.23\textwidth, keepaspectratio=true]{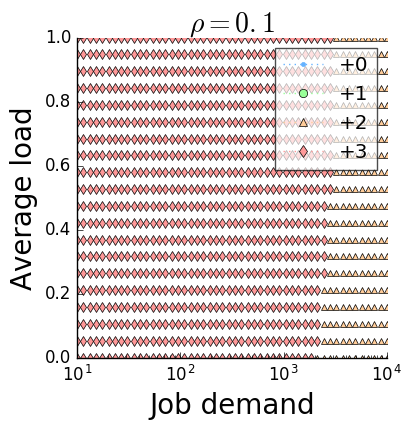}
  \end{subfigure}
  \begin{subfigure}
    \centering
    \includegraphics[width=.23\textwidth, keepaspectratio=true]{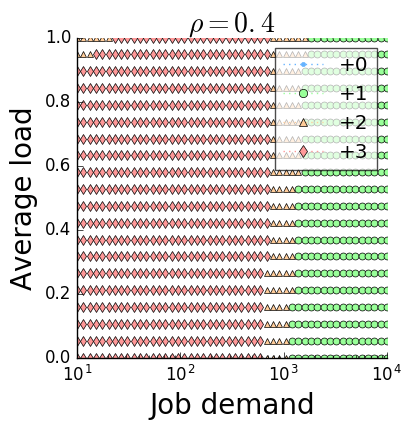}
  \end{subfigure}
  \begin{subfigure}
    \centering
    \includegraphics[width=.23\textwidth, keepaspectratio=true]{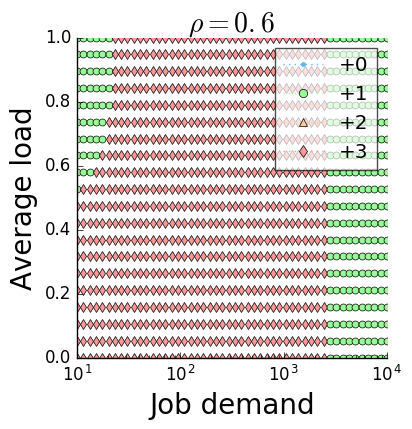}
  \end{subfigure}
  \begin{subfigure}
    \centering
    \includegraphics[width=.23\textwidth, keepaspectratio=true]{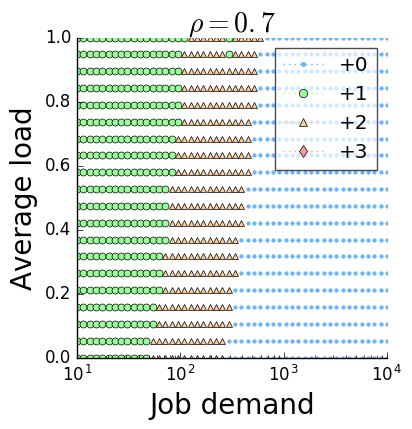}
  \end{subfigure}
  \caption{Scheduling policies learned by Deep Q-learning under varying offered load $\rho$ on the system. Given the demand of a job and the average load on its assigned cluster nodes, policy decides how many coded tasks to schedule for the job, e.g., +2 indicates 2 coded tasks.}
  \label{fig:plot_learned_policies}
\end{figure*}

Fig.~\ref{fig:plot_Qlearning_convergence} plots the evolution of Q-network's accuracy and the collected average reward over the learning episodes.
When $\rho$ is low ($=0.4$), jobs rarely wait in queue, hence job slowdowns are mainly due to runtime variability. This allows learning optimal Q-values, or equivalently settling down on a good scheduling policy, fairly quickly.
When $\rho$ is high ($=0.8$), however, jobs often wait in queue before getting scheduled, hence queueing times significantly affect the job slowdowns. Queueing times are determined by the complex system dynamics (not just runtime variability), thus they are inherently noisy.
This causes Q-learning to spend more time in exploration before being able to learn the optimal Q-values.
Spikes of high loss and low reward happen sporadically in any case, and this is because UCB pushes the scheduler to explore more rather than improving the Q-value estimates for the policy at hand.

Fig.~\ref{fig:plot_Esl_ET_wlearning} gives a performance comparison between the learned policy $\mathrm{Redundant\textnormal{-}small}$ and two naive policies: $\mathrm{Redundant\textnormal{-}all}$; scheduling all jobs with maximum redundancy, and $\mathrm{Redundant\textnormal{-}none}$; scheduling no job with redundancy.
When $\rho$ is high ($\rho > 0.6$), $\mathrm{Redundant\textnormal{-}all}$ overburdens the system with redundant tasks and drives it to instability.
$\mathrm{Redundant\textnormal{-}small}$ carefully employs redundancy so that stragglers are mitigated to some degree while the introduced redundancy does not overburden the system.
When $\rho$ is low, however, $\mathrm{Redundant\textnormal{-}small}$ performs worse than $\mathrm{Redundant\textnormal{-}all}$ (seen better in Fig.~\ref{fig:plot_slowdown_tail_wlearning}), in other words, $\mathrm{Redundant\textnormal{-}small}$ is sub-optimal in this case.
We explain why Q-learning might derive a sub-optimal policy for our problem in the following.

\vspace{0.5em}
\noindent
\textbf{What does Deep-RL learn?}
We here discuss the scheduling policies learned by Algorithm~\ref{alg:alg_q_learning}.
Recall that state inputs for scheduling a job are i) job demand, ii) average load on the cluster nodes that the jobs' tasks are assigned to.
Fig.~\ref{fig:plot_learned_policies} illustrates the learned policies for different values of offered load $\rho$.
Q-learning devises a natural strategy; learns to introduce gracefully less redundancy for larger values of job demand or $\rho$.
Average load on nodes assigned for job's tasks does not influence the scheduling decision much.

We previously observed that $\mathrm{Redundant\textnormal{-}small}$ performs worse than $\mathrm{Redundant\textnormal{-}all}$ when $\rho$ is low (recall Fig.~\ref{fig:plot_Esl_ET_wlearning}).
This is because RL learns to be rather conservative and schedule large jobs with less (or no) redundancy even when $\rho$ is quite low (see $\rho=0.1, 0.4$ in Fig.~\ref{fig:plot_learned_policies}).
We explain why Q-learning converges to this behavior as follows.
When a job with a large demand gets scheduled with redundancy, it occupies larger space in the system, and most likely for a significant duration. This causes the subsequent jobs to wait longer for enough resources to become available.
Given that job demands are heavy tailed, overwhelming majority of jobs behind a large job have short service time and their slowdown is highly aggravated by the waiting times.
Therefore, scheduling large jobs with redundancy might result in observing a long chain of jobs with high slowdown.
Having repeatedly observed poor performance as a result of scheduling large jobs with redundancy, the scheduler learns to be conservative in such situations. 


At this point, we got everything from Algorithm~\ref{alg:alg_q_learning} that we need to move on to the second stage of our scheduling policy design.
(Further improvement by Deep-RL techniques are usually achieved by fine tuning of the parameters, but that is not our approach.)


\section{Scheduling small jobs with redundancy}
\label{sec:sec_redsmall}
In this section, we use mathematical modeling to propose, study, and tune a scheduling policy, and show that the policy we devise with modeling and queueing analysis performs as good as the more complex policies derived by Deep-RL.

Deep-RL learns the following scheduling principles: \emph{right amount} of redundancy shall be introduced in executing \emph{small enough} jobs, and only when the offered load $\rho$ on the system is \emph{low enough}.
Building on these, we propose $\mathrm{Redundant\textnormal{-}small}$ policy that expands the arriving job with redundancy at a fixed rate of $r$ only if its demand is less than $d$.
When a job of $k$ tasks is scheduled with redundancy, it gets expanded into $\ceil{rk}$ tasks by $\ceil{rk} - k$ redundant tasks.
This adds redundancy in amounts proportional to jobs' initial number of tasks, which is fair in the sense that jobs with larger number of tasks are more likely to suffer from stragglers, hence get scheduled with more redundancy.
Note that Deep-RL scheduler in Sec.~\ref{sec:sec_learning} did not use a multiplicative rate $r$ to decide how much redundancy to add into jobs but directly tried to decide how many redundant tasks to add into each job. This allowed working with a discrete action space, which leads to more data efficient and easier implementation of Deep Q-learning.

Performance of $\mathrm{Redundant\textnormal{-}small}$ is shaped by its two parameters $r$ and $d$, and we will fix $r$ and optimize $d$.
Master-Worker compute system model that we adopted (as described in Sec.~\ref{sec:sec_sys_model}) is complex and formidable to study with an exact analysis. Therefore, we here present an approximate analysis and demonstrate that our approximation allows finding an accurate estimate of the optimal $d$.
We here continue to adopt the simplification given in Sec.~\ref{sec:sec_sys_model}, that is, tasks' requested resource capacity is fixed as $R = 1$. Derivations presented in the following can be extended to the case with random $R$ at the cost of more tedious expressions.

\vspace{0.5em}
\noindent
\textbf{Latency and Cost of job execution:}
We refer to the execution time of an arbitrary job as $\text{Latency}$, and the total resource time it consumes throughout its execution as $\text{Cost}$.
Recall that the random slowdown factor $S$ expands the service time of the tasks multiplicatively at runtime.
When scheduled with no redundancy, a job of $k$ tasks each with a minimum service time of $b$ completes once its slowest task finishes, hence its $\text{Latency} \sim b \times S_{k:k}$
and its $\text{Cost} \sim k \times b \times S$.
When scheduled together with $n-k$ (coded) redundant tasks, job will finish as soon as any $k$ of its $n$ tasks finish, hence its $\text{Latency} \sim  b \times S_{n:k}$ and its $\text{Cost} \sim b \times \left(\sum_{i=1}^{k-1} S_{n:i} + (n-k)S_{n:k}\right)$ \cite{MAMA:AktasPS17}.



By definition, the average system load, or equivalently the average load on any cluster node (recall that tasks of each job are dispatched to the least loaded set of nodes) is given by
\begin{equation}
  \rho = \frac{\lambda}{N C} \E[\text{Cost}].
\label{eq:eq_rho}
\end{equation}
Note that this expression holds for the Master-Worker system not only under $\mathrm{Redundant\textnormal{-}small}$ policy but under any work-conserving scheduling policy.
Executing a particular job with more redundancy always reduces its latency \cite{MAMA:AktasPS17}.
It has also been shown in \cite{MAMA:AktasPS17} that redundancy can also reduce the cost of a job, when the number of redundant tasks added into job is below a level and the runtime variability is heavy tailed beyond a level. 
This fact together with \eqref{eq:eq_rho} implies that executing jobs with redundancy can potentially decrease $\E[\text{Cost}]$ hence decrease $\rho$, which is likely to decrease the time jobs wait in the queue and reduce the overall slowdown experienced by the jobs (hence the good performance of $\mathrm{Redundant\textnormal{-}all}$ under low offered load, see Fig.~\ref{fig:plot_Esl_ET_wlearning}), or can increase $\E[\text{Cost}]$ hence increase $\rho$, which might further aggravate job slowdowns or even drive the system to instability (hence the poor performance or instability of $\mathrm{Redundant\textnormal{-}all}$ under high offered load).

Under $\mathrm{Redundant\textnormal{-}small}$ policy, a job of $k$ tasks with a service time of $b$ will be scheduled with redundancy only if its demand $D = kB \leq d$. By the law of total expectation,
\begin{equation}
\begin{split}
  \E[X] &= \E[X \mid D \leq d]\Pr\{D \leq d\} \\
    &\quad + \E[X \mid D > d]\left(1 - \Pr\{D \leq d\}\right),
\end{split}
\label{eq:eq_ESC_cond_sum}
\end{equation}
where $X$ is the placeholder for $\text{Latency}$ or $\text{Cost}$, and
\begin{equation}
\begin{split}
  \Pr\{D \leq d\} &= \Pr\{k B \leq d\} \\
    &= \E_k\left[\Pr\{B \leq d/k\}\right], \\
  \E[\text{Latency} \mid D > d] &= \E[B S_{k:k} \mid k B > d] \\
    &= \E_k\left[\E[S_{k:k}]\; \E[B \mid B > d/k]\right], \\
  \E[\text{Latency} \mid D \leq d] &= \E[B S_{n:k} \mid k B \leq d] \\
    &= \E_k\left[\E[S_{n:k}]\; \E[B \mid B \leq d/k]\right], \\
  \E[\text{Cost} \mid D > d] &= \E[k B S \mid k B > d] \\
    &= \E[S]\; \E_k\left[k\; \E[B \mid B > d/k]\right], \\
  \E[\text{Cost} \mid D \leq d] &= \E\left[B C_{n,k} \;\middle|\; k B \leq d \right] \\
    &= \E_k\left[\E[C_{n,k}]\; \E[B \mid B \leq d/k] \right].
\end{split}
\label{eq:eq_ESC_cond_parts}
\end{equation}
where $n = \ceil{kr}$ and $C_{n,k} = \sum_{i=1}^k S_{n:i} + (n-k)S_{n:k}$.
Using the results presented in \cite{MAMA:AktasPS17}, it is easy to derive
\begin{equation}
\begin{split}
  \E[S_{n:k}] &= \frac{\Gamma(n+1)}{\Gamma(n-k+1)} \frac{\Gamma(n-k+1-1/\alpha)}{\Gamma(n+1-1/\alpha)}, \\
  \E[C_{n,k}] &= \frac{n}{\alpha-1}\left(\alpha - \left(1 - k/n\right)\;\E[S_{n:k}] \right).
\end{split}
\label{eq:eq_ES_EC_nk}
\end{equation} Expected latency and cost can be computed using the expressions given above.
Another way to express them is
\begin{equation*}
\begin{split}
  & \E[\text{Latency}] = \E_k[\E[S_{k:k}]]\E[B] \\
  &\quad + \E_k\left[\E[S_{n:k} - S_{k:k}]\E[B \mid B \leq d/k]\Pr\{B \leq d/k\} \right], \\
  & \E[\text{Cost}] = \E[k]\E[B]\E[S] \\
  &\quad + \E_k\left[(\E[C_{n, k}] - k\E[S])\E[B \mid B \leq d/k]\Pr\{B \leq d/k\} \right].
\end{split}
\end{equation*}
These expressions better reflect the change in the latency and cost by increasing the demand threshold $d$ for selecting jobs to schedule with redundancy. In both expressions, the first term in the sum is equal to the baseline value of the expected latency or cost when no job is scheduled with redundancy (i.e., $d=0$).
For latency, the second term is always non-negative since $\E[S_{n:k}] \leq \E[S_{k:k}]$ for any $k$ and $n > k$, so redundancy always reduces the expected latency.
For cost, sign of the second term is given by the sign of $\E[C_{n, k}] - k\E[S]$, which can be either positive or negative depending on the values of $r$, $d$, $k$ and $n$, hence redundancy might increase the cost or even decrease it.
We elaborate on this and its consequences in system performance in the following.

We next derive an an asymptotic sufficient condition for a reduction in $\E[\text{Cost}]$ by employing $\mathrm{Redundant\textnormal{-}small}$ policy.
Gautschi's inequality \cite{NIST:Lozier03} gives us
\[ (1 - (k-1)/n)^{-1/\alpha} < \E[S_{n,k}] < (1 - (k+1)/n)^{-1/\alpha}. \]
Then, $\E[S_{n,k}]$ can be approximated as
\begin{equation}
  \E[S_{n,k}] \approx \left(1 - k/n\right)^{-1/\alpha}
\label{eq:eq_ESnk_approx}
\end{equation}
for $n > k$.
We numerically compute and report the relative error of the approximation given above with respect to the exact value of $\E[S_{n,k}]$ for $k = 5..20$, $n = k+1..2k$ and $\alpha = 2..10$ in Table~\ref{table:tb_ESnk_exact_vs_approx}.
The approximation is accurate (within 10\% relative error) even for small values of $k$.
\begin{table}[]
\begin{tabular}{|l|l|l|l|l|l|l|l|l|l|}
\cline{3-10}
\multicolumn{2}{}{} & \multicolumn{8}{|c|}{$\alpha$} \\ \hline
\multicolumn{1}{|c}{$k$} & \multicolumn{1}{|c|}{$n$} & 2 & 3 & 4 & 5 & 6 & 7 & 8 & 9 \\ \hline
\multirow{3}{*}{6} & 7 & 10.84 & 9.04 & 7.38 & 6.16 & 5.28 & 4.6 & 4.08 & 3.66 \\
  & 9 & 2.8 & 2.42 & 2.02 & 1.71 & 1.47 & 1.29 & 1.15 & 1.04 \\ 
  & 11 & 1.37 & 1.2 & 1.0 & 0.85 & 0.73 & 0.65 & 0.58 & 0.52 \\ 
\hline
\multirow{5}{*}{10} & 11 & 11.56 & 9.67 & 7.89 & 6.6 & 5.65 & 4.93 & 4.37 & 3.92 \\ 
  & 13 & 3.24 & 2.81 & 2.34 & 1.98 & 1.71 & 1.5 & 1.34 & 1.2 \\ 
  & 15 & 1.68 & 1.47 & 1.23 & 1.04 & 0.9 & 0.79 & 0.71 & 0.64 \\ 
  & 17 & 1.05 & 0.93 & 0.78 & 0.66 & 0.57 & 0.5 & 0.45 & 0.4 \\ 
  & 19 & 0.73 & 0.65 & 0.54 & 0.46 & 0.4 & 0.35 & 0.31 & 0.28 \\
\hline
\multirow{7}{*}{14} & 15 & 11.9 & 9.96 & 8.13 & 6.8 & 5.82 & 5.08 & 4.5 & 4.04 \\ 
  & 17 & 3.47 & 3.01 & 2.51 & 2.13 & 1.84 & 1.61 & 1.44 & 1.29 \\ 
  & 19 & 1.86 & 1.62 & 1.36 & 1.15 & 1.0 & 0.88 & 0.78 & 0.71 \\ 
  & 21 & 1.2 & 1.05 & 0.88 & 0.75 & 0.65 & 0.57 & 0.51 & 0.46 \\ 
  & 23 & 0.85 & 0.75 & 0.63 & 0.53 & 0.46 & 0.41 & 0.36 & 0.33 \\ 
  & 25 & 0.64 & 0.56 & 0.47 & 0.4 & 0.35 & 0.31 & 0.27 & 0.25 \\ 
  & 27 & 0.5 & 0.44 & 0.37 & 0.32 & 0.27 & 0.24 & 0.22 & 0.19 \\ 
\hline
\multirow{9}{*}{18} & 19 & 12.1 & 10.13 & 8.27 & 6.92 & 5.92 & 5.17 & 4.58 & 4.11 \\ 
  & 21 & 3.62 & 3.14 & 2.62 & 2.22 & 1.91 & 1.68 & 1.5 & 1.35 \\ 
  & 23 & 1.97 & 1.73 & 1.45 & 1.23 & 1.06 & 0.93 & 0.83 & 0.75 \\ 
  & 25 & 1.29 & 1.14 & 0.95 & 0.81 & 0.7 & 0.62 & 0.55 & 0.5 \\ 
  & 27 & 0.93 & 0.82 & 0.69 & 0.59 & 0.51 & 0.45 & 0.4 & 0.36 \\ 
  & 29 & 0.71 & 0.62 & 0.52 & 0.45 & 0.39 & 0.34 & 0.3 & 0.27 \\ 
  & 31 & 0.56 & 0.49 & 0.42 & 0.35 & 0.31 & 0.27 & 0.24 & 0.22 \\ 
  & 33 & 0.46 & 0.4 & 0.34 & 0.29 & 0.25 & 0.22 & 0.2 & 0.18 \\ 
  & 35 & 0.38 & 0.33 & 0.28 & 0.24 & 0.21 & 0.18 & 0.16 & 0.15 \\ \hline
\end{tabular}
\caption{Percentage error for the approximation given in \eqref{eq:eq_ESnk_approx} for $\E[S_{n,k}]$ for varying $k$, $n$ and $\alpha$.}
\label{table:tb_ESnk_exact_vs_approx}
\end{table}
Substituting \eqref{eq:eq_ESnk_approx} in \eqref{eq:eq_ES_EC_nk}, we obtain the following approximation for $\E[C_{n,k}]$
\begin{equation*}
\begin{split}
  \E[C_{n,k}] \approx \frac{n}{\alpha-1}\left(\alpha - \left(1 - k/n\right)^{1 - 1/\alpha} \right).
\end{split}
\end{equation*}
Approximating $n = \ceil{rk}$ by $rk$, we can write
\[ \E[C_{n,k}] \approx k f(\alpha, r), \]
for $f(\alpha, r) = \frac{r}{\alpha-1}\left(\alpha - \left(1 - 1/r\right)^{1 - 1/\alpha} \right)$.

Substituting this in $\E[\text{Cost} \mid D \leq d]$ given in \eqref{eq:eq_ESC_cond_parts}, we find
\begin{equation*}
\begin{split}
  \E[\text{Cost} \mid D \leq d] &= \E_k\left[f(\alpha, r)\; k\;\E[B \mid B \leq d/k] \right] \\
  &\approx f(\alpha, r)\; \E[k B \mid k B \leq d].
\end{split}
\end{equation*}
Finally substituting this together with \eqref{eq:eq_ESC_cond_parts} in \eqref{eq:eq_ESC_cond_sum} yields
\begin{equation*}
\begin{split}
  \E[\text{Cost}] \approx\; &\E[k]\E[D]\E[S] + \Pr\{kB \leq d\} \\
  &\times \E[kB \mid kB \leq d]\left(f(\alpha, r) - \E[S]\right)
\end{split}
\end{equation*}
First term in this sum is the average cost with no redundancy.
Average cost is reduced by $\mathrm{Redundant\textnormal{-}small}$ if and only if $f(\alpha, r) - \E[S] \lesssim 0$, which can be rewritten by substituting in $\E[S] = \alpha/(\alpha - 1)$ as
\begin{equation}
  r \lesssim \left(1 - \alpha^{-\alpha} \right)^{-1}
\label{eq:eq_approx_cond_toreduce_EC}
\end{equation}
for $\alpha > 1$.
Condition \eqref{eq:eq_approx_cond_toreduce_EC} only depends on the job expansion rate $r$ and tail index $\alpha$ of runtime slowdown factor ($S$), but depends neither on the threshold in job demand for stopping to add redundancy ($d$) nor on the distribution of the number of tasks ($k$) and task service times ($B$).
This suggests an interesting characteristic of $\mathrm{Redundant\textnormal{-}small}$, that is, expanding all jobs with coded redundancy will not only reduce latency but also reduce average cost as long as the expansion rate $r$ is kept below a threshold, which is determined solely by the runtime variability.
As can be seen in \eqref{eq:eq_rho}, reduced average cost will translate into reduced average load $\rho$ exerted on the system, and this reduction in load will most likely reduce the time that jobs spend waiting in the queue for enough resources to become available in the cluster in order to start execution.

In Sec.~\ref{sec:sec_sys_model}, we set the tail index of runtime variability as $\alpha = 3$, and for this case \eqref{eq:eq_approx_cond_toreduce_EC} gives us the condition $r \lesssim 1.038$. Expanding jobs at a rate as low as $1.038$ implies scheduling only the jobs of \emph{many} tasks with redundancy.
This implies that average cost of job execution can be reduced with $\mathrm{Redundant\textnormal{-}small}$ only if the jobs that run at large scale get scheduled with redundancy.
This is a natural consequence of the fact that the impact of stragglers on the latency grows with the scale of job execution, hence execution with coded redundancy is more effective at larger scale \cite{MAMA:AktasPS17}.
Thus, average cost of job execution (consequently $\rho$) can be reduced only if redundant tasks, which can potentially increase the cost, are employed for jobs that would benefit the most from them.

Note that \eqref{eq:eq_approx_cond_toreduce_EC} is only a sufficient condition reduce $\E[\text{Cost}]$ and consequently $\rho$. As will be shown in the remainder of this section, it is possible to reduce job slowdowns with $\mathrm{Redundant\textnormal{-}small}$ by \emph{carefully} adjusting the value of $d$, even at the expense of aggravated $\E[\text{Cost}]$ and $\rho$.

\begin{figure*}[t]
  \centering
  \begin{subfigure}
    \centering
    \includegraphics[width=.185\textwidth, keepaspectratio=true]{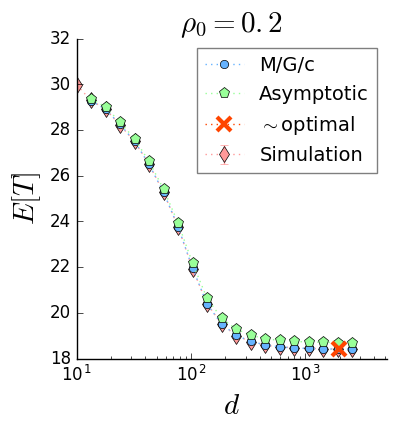}
  \end{subfigure}
  \begin{subfigure}
    \centering
    \includegraphics[width=.185\textwidth, keepaspectratio=true]{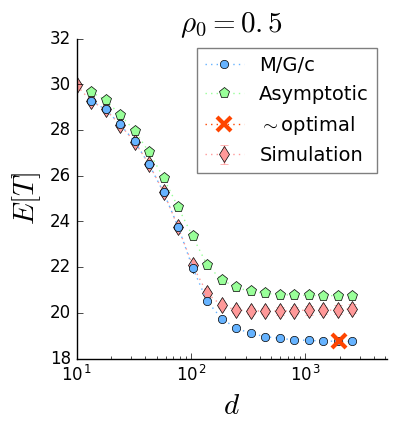}
  \end{subfigure}
  \begin{subfigure}
    \centering
    \includegraphics[width=.185\textwidth, keepaspectratio=true]{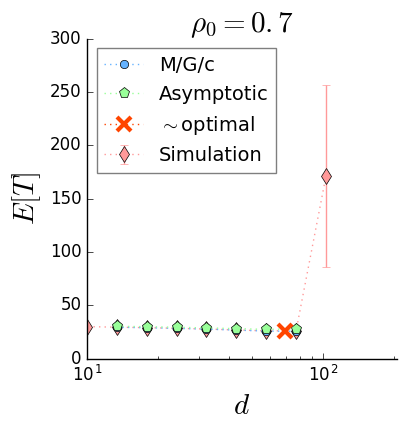}
  \end{subfigure}
  \begin{subfigure}
    \centering
    \includegraphics[width=.185\textwidth, keepaspectratio=true]{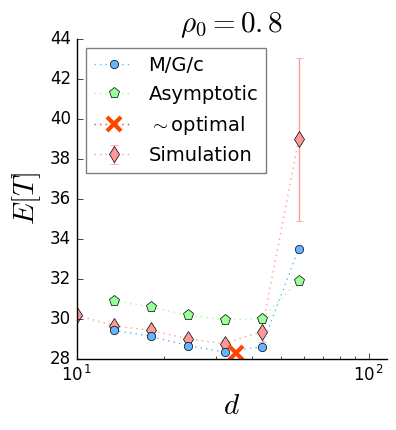}
  \end{subfigure}
  \begin{subfigure}
    \centering
    \includegraphics[width=.185\textwidth, keepaspectratio=true]{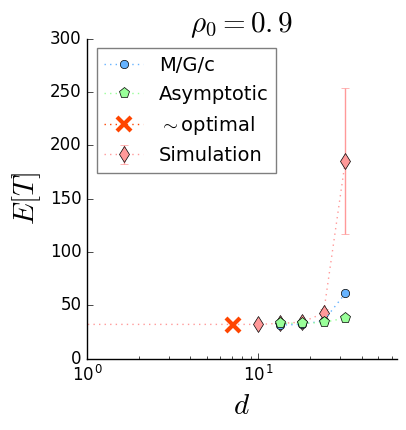}
  \end{subfigure}
  \caption{Average system response time $\E[T]$ under $\mathrm{Redundant\textnormal{-}small}$ policy with job expansion rate $r=2$ for varying levels of offered load $\rho_0$. $M/G/c$ and asymptotic refer to the values estimated by \eqref{eq:eq_ET_approx} in Claim~\ref{claim_ET} and its equivalent at large scale limit.
  Red-cross shows the optimal $d^*$ that minimizes $\myapprox\E[T]$ given in \eqref{eq:eq_ET_approx}, and $d^* < 10$ found for $\rho_0=0.9$ implies scheduling no job with redundancy.}
  \label{fig:plot_redsmall_ET}
\end{figure*}

\vspace{0.5em}
\noindent
\textbf{$\mathbf{M/G/c}$ approximation.}
We here explain how to approximate the Master-Worker compute system that we consider (described in Sec.~\ref{sec:sec_sys_model}) as an $M/G/c$ queue.
An $M/G/c$ queue is a simplified Master-Worker system; it denotes a first-in first-out queue receiving Poisson single-task job arrivals and feeding $c$ servers by dispatching the job at the head of the queue as soon as a server becomes idle.
Each server can serve a single job at a time with service times i.i.d. across different jobs and servers.
It is a well studied model and numerous approximations are available for its average response time \cite{TelephoneCallCenters:GansKM03}.

One can analytically analyze multi-server queueing systems if each job takes up a fixed space in the system while spending a random amount of time in service.
This is not the case in the Master-Worker system, but our idea is to make an approximation as follows.
We first assume that each job consumes a fixed capacity of $\sigma$ per unit time, as is the case with an $M/G/c$ queue.
Total capacity available per unit time in our system is $NC$.
We can think of dividing it into channels of capacity $\sigma$, where each channel is assigned to a different job.
Then, it becomes natural to treat the system as an $M/G/c$ queue with $NC/\sigma$ servers.
On average, an arbitrary job spends $\E[\text{Latency}]$ of time in service and consumes $\E[\text{Cost}]$ of capacity throughout its service, hence it consumes $\E[\text{Cost}]/\E[\text{Latency}]$ capacity per unit time on average. This is an unbiased estimator of $\sigma$ and we use it as an approximation for $\sigma$.
\begin{approximation}
  Master-Worker system described in Sec.~\ref{sec:sec_sys_model} is approximated as an $M/G/c$ queue with $NC\; \E[\text{Latency}]/\E[\text{Cost}]$ servers and with job service times distributed as $\text{Latency}$.
\label{approx_MGc}
\end{approximation}
The expression given for the number of servers $c$ in Approximation~\ref{approx_MGc} can be non-integer. As will be seen shortly, we circumvent this by transforming the expressions so that they work with non-integer $c$.
The most well known approximation for the average response time in $M/G/c$ queue is given by adjusting the average waiting time in its Markovian counterpart, $M/M/c$ queue, as
\begin{equation}
  \E[T_{M/G/c}] \approx \E[X] + \frac{\mathcal{C}^2 + 1}{2} \E[W_{M/M/c}],
\label{eq:eq_ET_MGc}
\end{equation}
where $X$ is the service time distribution and $\mathcal{C}$ is its coefficient of variation. We know
\[ \E[W_{M/M/c}] = \Pr\{\mathrm{Queueing}\} \frac{\rho}{\lambda(1 - \rho)}, \]
where $\rho = \lambda\E[X]/c$ denotes the average load on a server. $\Pr\{\mathrm{Queueing}\}$ denotes the probability that an arriving job waits in the queue before starting service and is given by
\[ \Pr\{\mathrm{Queueing}\} = \left(1 + (1-\rho) \frac{c!}{(c\rho)^c} \sum_{i=0}^{c-1} \frac{(c\rho)^i}{i!} \right)^{-1}, \]
which is known as Erlang's C formula.
The above exponential sum can be written in terms of the incomplete Gamma function $\Gamma(a, x) = \int_{x}^{\infty} u^{a-1} e^{-u} du$, so we have
\begin{equation}
  \Pr\{\mathrm{Queueing}\} = \left(1 + (1-\rho) \frac{c\;e^{c\rho}}{(c\rho)^c}\Gamma(c, c\rho) \right)^{-1}.
\label{eq:eq_Pr_Queueing}
\end{equation}
The form given above now is defined for non-integer $c$.
At large scale limit, i.e., keeping $\rho$ fixed while taking $c\rho \to \infty$,
\[ \lim_{c\rho \to \infty} \Gamma(c, c\rho) = (c\rho)^{c-1} e^{-c\rho}. \]
Substituting this into \eqref{eq:eq_Pr_Queueing}, we get
\begin{equation}
  \lim_{c\rho \to \infty} \Pr\{\mathrm{Queueing}\} = \left(1 + (1-\rho)/\rho\right)^{-1} = \rho.
\label{eq:eq_Pr_queueing_largescalelimit}
\end{equation}

We next use this $M/G/c$ queue approximation to find an approximation for the average response time in our Master-Worker system.
Note that the approximate expression in \eqref{eq:eq_ET_approx} requires the coefficient of variation $\mathcal{C}$ for the service time distribution for the Approximation~\ref{approx_MGc}, which is given as
\[ \mathcal{C}^2 = \E[\text{Latency}^2]/\E[\text{Latency}]^2 - 1. \]
First moment of $\text{Latency}$ has been previously derived using the law of total expectation and the expressions given in \cite{MAMA:AktasPS17} (recall \eqref{eq:eq_ESC_cond_sum}, \eqref{eq:eq_ESC_cond_parts} and \eqref{eq:eq_ES_EC_nk}). Second moment is derived exactly the same way and omitted here because of the space constraint.
\begin{claim}
  Average response time $\E[T]$ in the Master-Worker system described in Sec.~\ref{sec:sec_sys_model} is approximated as
  \begin{equation}
  \begin{split}
    \E[T] \approx\; &\E\ \\
    &+ \frac{\E[\text{Latency}^2]}{2\E[\text{Latency}]^2} \Pr\{\mathrm{Queueing}\} \frac{\rho}{\lambda(1 - \rho)}.
  \end{split}
  \label{eq:eq_ET_approx}
  \end{equation}
  Under $\mathrm{Redundant\textnormal{-}small}$ policy, $\E[\text{Latency}]$ is given by \eqref{eq:eq_ESC_cond_sum} (and so $\E[\text{Latency}^2]$ is given similarly), $\rho$ is given by \eqref{eq:eq_rho}, and $\Pr\{\mathrm{Queueing}\}$ is given by \eqref{eq:eq_Pr_Queueing} and by \eqref{eq:eq_Pr_queueing_largescalelimit} at large scale limit.
\label{claim_ET}
\end{claim}

Fig.~\ref{fig:plot_redsmall_ET} gives a comparison between the simulated values of average response time $\E[T]$ under $\mathrm{Redundant\textnormal{-}small}$ policy and the values estimated by the approximate expression \eqref{eq:eq_ET_approx}.
Values estimated by the $M/G/c$ queue approximation overall follow the simulated ones fairly closely.
When the offered load $\rho_0$ (baseline load when no job is scheduled with redundancy) on the system is low ($\rho_0 \leq 0.5$), scheduling even very large jobs with redundancy ($d \to \infty$) does not hurt performance but further reduces the job slowdowns (recall the good performance of $\mathrm{Redundant\textnormal{-}all}$ under low offered load in Fig.~\ref{fig:plot_Esl_ET_wlearning}).
When $\rho_0$ is high, scheduling with redundancy reduces $\E[T]$ until $d$ reaches a threshold, beyond which increasing $d$ hurts performance ($\rho_0 = 0.6$), and might even drive the system to instability depending on $\rho_0$ (see the plots for $\rho_0 \geq 0.7$).

\begin{figure}[h]
  \centering
  \begin{subfigure}
    \centering
    \includegraphics[width=.23\textwidth, keepaspectratio=true]{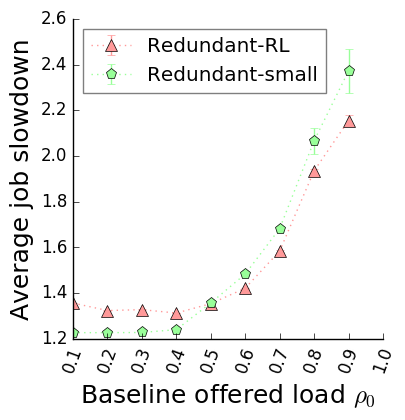}
  \end{subfigure}
  \begin{subfigure}
    \centering
    \includegraphics[width=.23\textwidth, keepaspectratio=true]{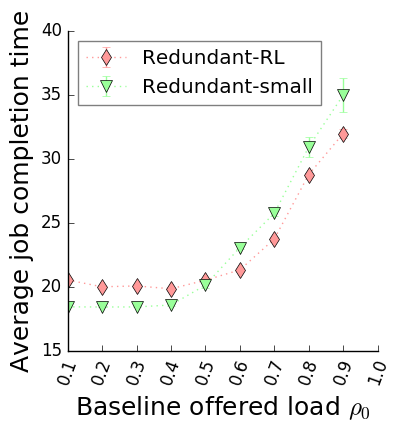}
  \end{subfigure}
  \caption{Average job slowdown and completion times for $\mathrm{Redundant\textnormal{-}RL}$ (RL) and $\mathrm{Redundant\textnormal{-}small}$ (Red-small) with job expansion rate $r=2$ and redundancy threshold $d$ that is analytically optimized using the approximation given in Claim~\ref{claim_ET}.}
  \label{fig:fig_redsmall_vs_DRL}
\end{figure}

\vspace{0.5em}
\noindent
\textbf{$\mathbf{Redundant\textnormal{-}small}$ vs.\ $\mathbf{Redundant\textnormal{-}RL}$:}
Most importantly, approximate expression \eqref{eq:eq_ET_approx} is able to accurately estimate the trajectory of decrease or increase in the average system response time $\E[T]$ as $d$ increases goes from $0$ to $\infty$. This enables us to accurately estimate the optimal demand threshold $d^*$ for selecting jobs to schedule with redundancy (shown with red-cross in Fig.~\ref{fig:plot_redsmall_ET}) that minimizes $\E[T]$.
A performance comparison is shown in Fig.~\ref{fig:fig_redsmall_vs_DRL} between the policies learned by Deep-RL and the $\mathrm{Redundant\textnormal{-}small}$ policy with approximately computed optimal $d^*$.
$\mathrm{Redundant\textnormal{-}small}$ overall performs as good as the more complex policies learned by Deep-RL.
When the offered load $\rho_0$ is low, approximation \eqref{eq:eq_ET_approx} guides us to set $d^*$ to a large value hence $\mathrm{Redundant\textnormal{-}small}$ acts as $\mathrm{Redundant\textnormal{-}all}$. As $\rho_0$ gets higher, approximation guides us to decrease $d^*$ gradually, similar to the behavior learned by Deep-RL (recall Sec.~\ref{sec:sec_learning}). When $\rho_0$ is large (close to 1), approximation tells us to set $d^*$ to zero hence no job is scheduled with redundancy as in $\mathrm{Redundant\textnormal{-}none}$ in that case.

\section{Straggler Relaunch}
\label{sec:sec_relaunch}
After a job starts execution, waiting for reasonably long, and treating the remaining tasks that did not complete by then as possible stragglers and relaunching them has been shown to be effective in mitigating the impact of stragglers both in practice \cite{ImprovingMapReducePerformance:ZahariaKJ08} and theory \cite{IFIP:AktasPS17}.
Effectiveness of straggler relaunch relies on the tail heaviness of the runtime variability because relaunching is a choice of canceling the work that is already completed in order to get possibly lucky and execute the fresh replacement copies much faster.
Heavy tailed nature of the runtime slowdown implies that if we wait for a reasonably large $\Delta$ amount of time after a job starts execution and find the job still not completed by then, we expect only a few tasks to be straggling and such tasks are expected to take at least $\Delta$ more to complete on average.
These two observations imply that if a fresh copy is launched at time $\Delta$ for each straggling task, then each fresh copy is likely to complete before the corresponding old copy, thus help to complete the job at hand earlier (reducing latency) and release the resources occupied by the straggling tasks faster (reducing cost).

Straggler relaunch has been shown to reduce the latency and cost of job execution when the runtime slowdown is heavy tailed beyond a level, and relaunching is performed at the right time \cite{IFIP:AktasPS17}.
We here adopt the same latency and cost definitions introduced in Sec.~\ref{sec:sec_redsmall}; $\text{Latency}$ and $\text{Cost}$ respectively refer to the time an arbitrary job spends in service at the cluster nodes and the total resource time it consumes throughout its execution.
We also adopt the assumption of \cite{IFIP:AktasPS17} that cancelling and relaunching tasks take place instantly hence do not incur any additional delay.
The set of tasks that will be relaunched for a job is decided by the amount of time $\Delta$ we wait before relaunching the remaining tasks, which we refer to as the \emph{relaunch time} for the job.
Relaunching tasks untimely might hurt performance; late relaunch leads to delayed cancellation of the stragglers or early relaunch causes cancelling the non-straggler tasks together with stragglers.
Optimal relaunch time that minimizes the latency (and cost) of job execution is found in \cite{IFIP:AktasPS17} to be approximately given as
\begin{equation}
  \Delta^* \approx b\sqrt{\frac{k!\Gamma(1-1/\alpha)}{\Gamma(k+1-1/\alpha)}},
\label{eq:eq_approx_optd_wrelaunch}
\end{equation}
which says that relaunch time for a job shall be set to the product of the minimum service time ($b$) of its tasks and a factor $w$ that is determined by the number of tasks within the job ($k$) and the tail heaviness of the runtime slowdown factor ($\alpha$).
Inspired by this result, we here consider the $\mathrm{Straggler\textnormal{-}relaunch}$ policy for the Master-Worker compute system, which assigns a relaunch time $b \times w$ for each arriving job and relaunches the remaining tasks in the job as soon as its relaunch time expires.

In Sec.~\ref{sec:sec_redsmall}, we proposed an $M/G/c$ queue approximation for the Master-Worker system (as stated in Approximation~\ref{approx_MGc}), and showed that it is fairly accurate in estimating the average system response time $\E[T]$ under $\mathrm{Redundant\textnormal{-}small}$ policy.
This approximation can be extended for the system with $\mathrm{Straggler\textnormal{-}relaunch}$ policy. Then, $\E[T]$ in the system under $\mathrm{Straggler\textnormal{-}relaunch}$ policy can be estimated using the approximate expression \eqref{eq:eq_ET_approx} previously given in Claim~\ref{claim_ET}, by replacing the moments of latency and cost with the ones that correspond to the relaunch policy.
We next describe how to derive $\E[\text{Latency}]$, $\E[\text{Latency}^2]$ and $\E[\text{Cost}]$ for job execution under $\mathrm{Straggler\textnormal{-}relaunch}$ policy.
As in Sec.~\ref{sec:sec_redsmall}, we here continue to assume that tasks' requested resource capacity is fixed as $R = 1$, and the derivations can be extended to the case with random $R$ by simply adding a layer of expectation with respect to $R$ in the expression given for $\E[\text{Cost}]$ in \eqref{eq:eq_ES_ES2_EC_relaunch} that is soon to be presented in the following.
Results presented in \cite{IFIP:AktasPS17} imply for a job of $k$ tasks with a minimum service time of $b$ that is executed with a relaunch time factor of $w$,
\begin{equation*}
\begin{split}
  \E[\text{Latency}_{k, b}] &= b\;w(1-q^k) + b\;\frac{\Gamma(k+1)\Gamma(1-1/\alpha)}{\Gamma(k+1-1/\alpha)} \\
    &\quad \times \left((1/w-1) I(1-q; 1-1/\alpha, k) + 1\right), \\
  \E[\text{Cost}_{k, b}] &= b\;k\frac{\alpha}{\alpha-1}\left((1-q)(1-w) + 1\right),
\end{split}
\end{equation*}
where $q = 1 - (1/w)^{\alpha}$, and recall the slowdown factor $S \sim \mathrm{Pareto}(1, \alpha)$.
We found the second moment of latency as
\begin{equation*}
\begin{split}
  \E[\text{Latency}^{2}&_{k, b}] = b^2\biggl(w^2(1 - q^k) + f(2)\frac{\Gamma(1-2/\alpha)}{\Gamma(1-1/\alpha)} \\
  &+ 2 w f(1)(1-q)^{1/\alpha}I(1-q; 1-1/\alpha, k) \\
  &+ (1 - w^2)f(2)(1-q)^{2/\alpha}I(1-q; 1-2/\alpha, k)\biggr).
\end{split}
\end{equation*}
for $f(i) = \Gamma(k+1)\Gamma(1-i/\alpha)/\Gamma(k+1-i/\alpha)$. 
Derivation of this is lengthy and omitted here for brevity.
Finally, we can express the first and/or second moments of latency and cost for an \emph{arbitrary} job as
\begin{equation}
\begin{split}
  \E[\text{Latency}] &= \E_k\left[\E_B\left[\E\left[\text{Latency}_{k, B} \right]\right]\right], \\
  \E[\text{Latency}^2] &= \E_k\left[\E_B\left[\E\left[\text{Latency}^2_{k, B} \right]\right]\right], \\
  \E[\text{Cost}] &= \E_k\left[\E_B\left[\E\left[\text{Cost}_{k, B} \right]\right]\right].
\end{split}
\label{eq:eq_ES_ES2_EC_relaunch}
\end{equation}

\begin{figure*}[t]
  \centering
  \begin{subfigure}
    \centering
    \includegraphics[width=.185\textwidth, keepaspectratio=true]{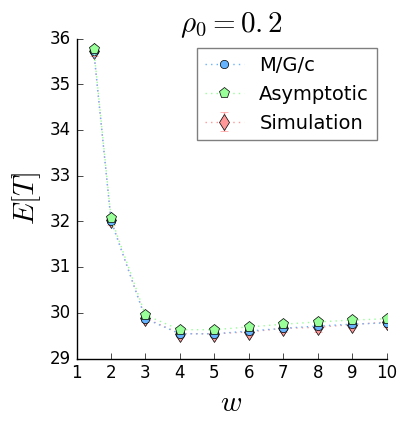}
  \end{subfigure}
  \begin{subfigure}
    \centering
    \includegraphics[width=.185\textwidth, keepaspectratio=true]{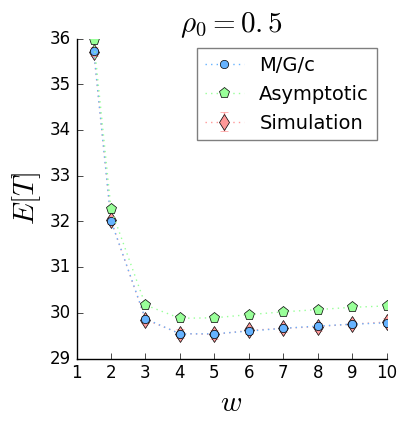}
  \end{subfigure}
  \begin{subfigure}
    \centering
    \includegraphics[width=.185\textwidth, keepaspectratio=true]{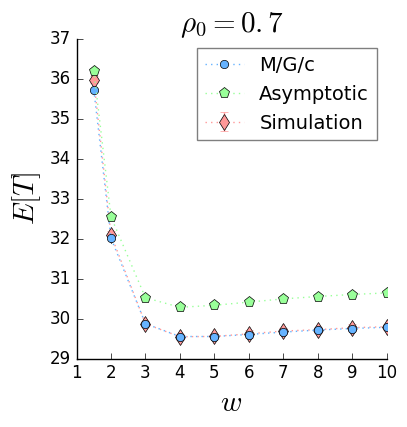}
  \end{subfigure}
  \begin{subfigure}
    \centering
    \includegraphics[width=.185\textwidth, keepaspectratio=true]{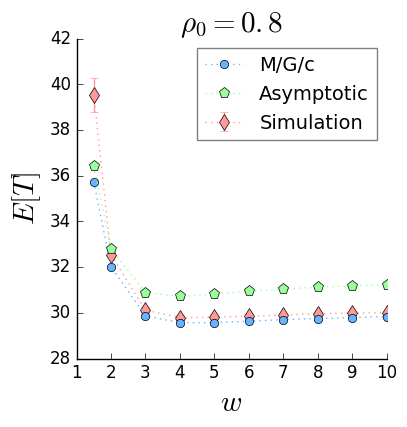}
  \end{subfigure}
  \begin{subfigure}
    \centering
    \includegraphics[width=.185\textwidth, keepaspectratio=true]{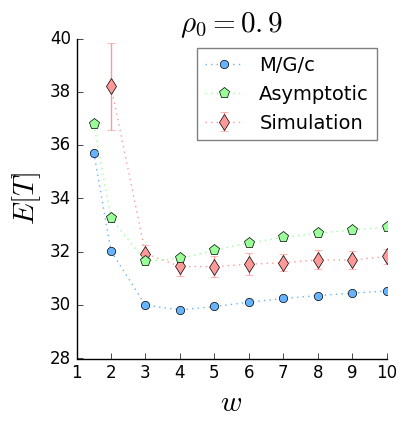}
  \end{subfigure}
  \caption{Average response time of system with $\mathrm{Straggler\textnormal{-}relaunch}$ policy under varying offered load $\rho_0$.
  $M/G/c$ refers to the values estimated by substituting \eqref{eq:eq_ES_ES2_EC_relaunch} in the approximate expression \eqref{eq:eq_ET_approx} given in Claim~\ref{claim_ET}, and asymptotic refers to the same approximation at large scale limit.
  When $\rho_0 = 0.9$ and $w = 1.5$, straggler relaunch drives system to instability, which is skipped to be able to show the values for $w > 1.5$.}
  \label{fig:plot_relaunch_ET}
\end{figure*}

To evaluate the accuracy of the $M/G/c$ approximation, we simulated the cluster with $\mathrm{Straggler\textnormal{-}relaunch}$ policy by fixing its relaunch time factor $w$ to the same value for each arriving job. For instance when $w=2$, relaunch time for an arriving job is set to twice of the job's minimum service time.
Fig.~\ref{fig:plot_relaunch_ET} gives a comparison between the simulated values for the average response time $\E[T]$ and the values estimated by the $M/G/c$ approximation.
Note that system in this case is configured as described in Sec.~\ref{sec:sec_sys_model}.
As can be seen in Fig.~\ref{fig:plot_relaunch_ET}, and in several other simulation results that we generated for other system configurations (but excluded their plots here because of the space constraint), we found that $M/G/c$ approximation is able to yield accurate estimates for $\E[T]$.

\begin{figure}[h]
  \centering
  \begin{subfigure}
    \centering
    \includegraphics[width=.23\textwidth, keepaspectratio=true]{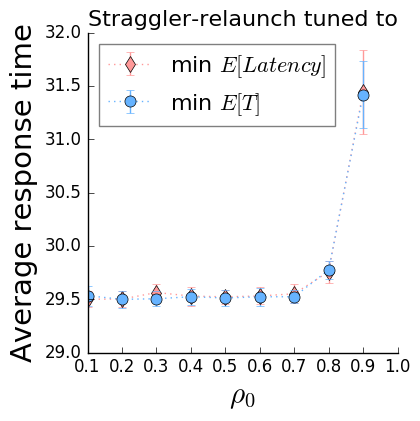}
  \end{subfigure}
  \begin{subfigure}
    \centering
    \includegraphics[width=.23\textwidth, keepaspectratio=true]{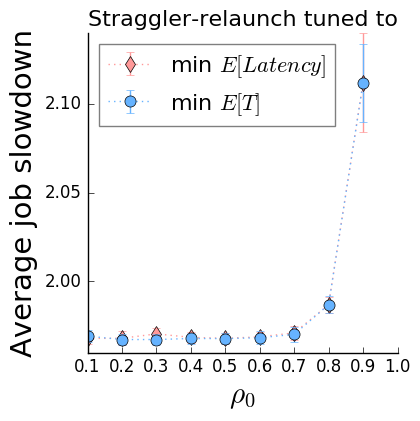}
  \end{subfigure}
  \caption{System performance under $\mathrm{Straggler\textnormal{-}relaunch}$ by setting the value of the multiplicative relaunch time factor $w$ differently for each job in order to minimize its $\E[\text{Latency}]$, or by fixing it for all jobs and setting its value in order to minimize $\E[T]$.}
  \label{fig:fig_relaunch_optimize_wrtES_vs_wrtET}
\end{figure}

Performance of $\mathrm{Straggler\textnormal{-}relaunch}$ can be optimized by tuning its multiplicative relaunch time factor $w$.
Recall that the simulated values in Fig.~\ref{fig:plot_relaunch_ET} are generated by fixing $w$ to the same value for all jobs, and a clear increase can be seen in $\E[T]$ as $w$ increases beyond a value $\myapprox 4$.
There are two ways to optimize the system performance:
1) Fix $w$ for all jobs and set it to a value that minimizes $\E[T]$,
2) Set $w$ differently for each arriving job in order to minimize the job's latency, which implies also minimizing its cost.
The first approach can be done using the approximate expression for $\E[T]$ given in \eqref{eq:eq_ET_approx}.
The second approach can be done by numerically computing the optimal $w$ for each arriving job using the exact expression of $\E[\text{Latency}]$ given in \eqref{eq:eq_ES_ES2_EC_relaunch}, or by directly using the approximate optimal relaunch time given in \eqref{eq:eq_approx_optd_wrelaunch}.
Fig.~\ref{fig:fig_relaunch_optimize_wrtES_vs_wrtET} plots the system performance of $\mathrm{Straggler\textnormal{-}relaunch}$ policy by optimizing the value of $w$ using the first or the second approach described above.
This plot, together with others that we generated for different system configurations but omitted here for brevity, tells us that there is almost no difference between these two ways of performance tuning in terms of $\E[T]$ and average job slowdown achieved by the system.

\begin{figure}[h]
  \centering
  \begin{subfigure}
    \centering
    \includegraphics[width=.23\textwidth, keepaspectratio=true]{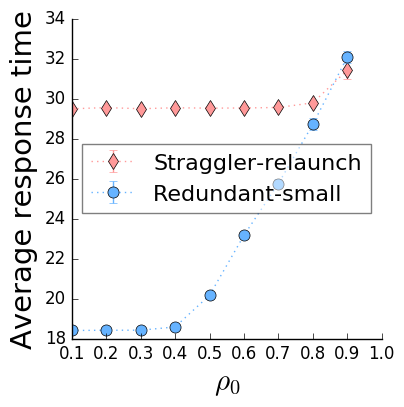}
  \end{subfigure}
  \begin{subfigure}
    \centering
    \includegraphics[width=.23\textwidth, keepaspectratio=true]{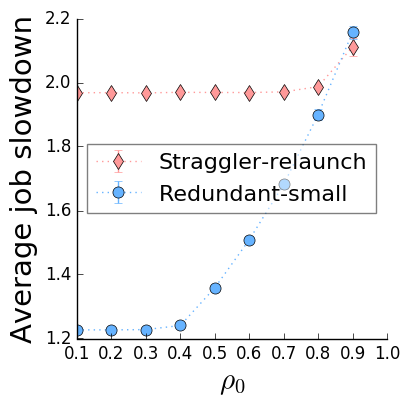}
  \end{subfigure}
  \caption{System performance comparison under optimized $\mathrm{Redundant\textnormal{-}small}$ and $\mathrm{Straggler\textnormal{-}relaunch}$ policies.}
  \label{fig:fig_StragglerRelaunch_vs_RedSmall}
\end{figure}

We finally compare the performance of $\mathrm{Redundant\textnormal{-}small}$ and $\mathrm{Straggler\textnormal{-}relaunch}$, while tuning both policies to minimize $\E[T]$ using the corresponding approximate expression obtained by the $M/G/c$ approximation.
Fig.~\ref{fig:fig_StragglerRelaunch_vs_RedSmall} shows the system performance under either policy for varying levels of offered load on the system.
As can be seen in the plot (and others that we generated for different system configurations but omitted here for brevity), when the offered load is low or moderate ($\rho_0 \leq 0.8$), execution with redundancy yields significantly lower job slowdowns compared to employing straggler relaunch instead.
When the offered load gets very high, $\mathrm{Straggler\textnormal{-}relaunch}$ starts to slightly outperform $\mathrm{Redundant\textnormal{-}small}$. This can be explained as follows. The (optimized) straggler relaunch policy only reduces the latency and cost of each arriving job, hence does not overburden the system with increased cost while mitigating the impact of stragglers to some degree under any offered load.
On the other hand, executing jobs with redundancy is certain to decrease latency while also very likely to increase the cost. 
Increased cost is acceptable as long as the offered load is below some level since it would not cause much increase in the duration jobs wait in the queue for resources to become available, thus $\mathrm{Redundant\textnormal{-}small}$ employs redundancy aggressively when the offered load is low or moderate and mitigates the impact of stragglers much more effectively than $\mathrm{Straggler\textnormal{-}relaunch}$.
$\mathrm{Redundant\textnormal{-}small}$ regulates the overall increase in cost by lowering the job demand threshold $d$ and selecting only fewer and smaller jobs to schedule with redundancy as the offered load gets higher.
If the offered load is really high, then $\mathrm{Redundant\textnormal{-}small}$ chooses to add no redundancy to any job in order not to aggravate the queueing times, thus not mitigating the impact of stragglers at all and performing slightly worse than $\mathrm{Straggler\textnormal{-}relaunch}$.
We found using simulations and the approximate expression we presented for $\E[T]$ that advantage of $\mathrm{Straggler\textnormal{-}relaunch}$ over $\mathrm{Redundant\textnormal{-}small}$ at very high offered load becomes more apparent under \emph{heavier tailed} task service times.

\newpage
\section{On the shortcomings of our system model \\ and approximate analysis}
\label{sec:sec_shortcomings}
\noindent
\textbf{Runtime slowdown model:}
We adopted the runtime variability model that was shown to support the experimental evidence in \cite{MASCOTS:GardnerHS16}.
However, this model is still optimistic; it ignores the effect of added redundant tasks on the chance and impact of straggling experienced by the tasks.
Despite much recent effort in implementing advanced resource usage isolation between different workloads \cite{BorgOmegaKubernetes:Burns16, Mesos:HindmanKZ2011}, contention at the resources located on the nodes (e.g., CPU, memory, I/O bus) or within the intra-cluster connection infrastructure (e.g., network links and switches) is still the primary cause of runtime variability \cite{TailAtScale:DeanB13, StragglerRootCauseAnalysisInDatacenters:OuyangGY16, RootCauseAnalysisOfStragglersInBigDataSystem:ZhouLY18}.
Redundant tasks added into the system is most likely to increase the existing contention for cluster resources, hence aggravating the runtime variability.
In addition, we assume cancelling redundant tasks (or relaunching the original tasks) takes place instantly, as commonly assumed in the existing theory research on systems with redundancy, but it will surely take some time in practice and can be significant at large scale.
These are not captured by the widely used model for systems with redundancy, which we also adopt here, and that is why the results showing the effectiveness of redundancy (or straggler relaunch) in reducing job slowdowns is an optimistic guess for how they would perform in practice.

It is reasonable to expect the runtime variability to get worse as the load exerted on the system gets higher.
One possible way to capture this in the model would be adopting a runtime slowdown factor that is shaped by the load exerted on the system.
For instance, in this paper we modeled the slowdown factor as a random variable $S$ that is distributed as a $\mathrm{Pareto}$ with a minimum value of $1$ and tail index $\alpha$. As the load on the system increases, $S$ is expected to get ``stochastically bigger''. To expand it a bit more, let slowdown factor be $S$ when the system load is $\rho$ and be $S^\prime$ when the load is $\rho^\prime > \rho$, then we expect $\Pr\{S^\prime > s\} \geq \Pr\{S > s\}$ for all $s$.
In the $\mathrm{Pareto}$ slowdown model, this can be modeled with a reduction in the tail index $\alpha$ (an increase in tail heaviness), while keeping its minimum value fixed to $1$ since it is always possible for tasks to finish execution without straggling.
Reduction in $\alpha$ can be modeled by defining it as a decreasing function of $\rho$.
Adopting such a model for $S$ would turn the expressions presented for system load \eqref{eq:eq_rho} and average system response time \eqref{eq:eq_ET_approx} into a recurrence relation in terms of $\rho$, which can be solved numerically.
It would also be interesting to see whether Deep-RL learns a different policy (than what we observed in Sec.~\ref{sec:sec_learning}) for scheduling with redundancy under this new slowdown model, which we leave as a future work.

In our system model, we assume service times of the tasks within each arriving job are known at the scheduling time.
Both scheduling policies $\mathrm{Redundant\textnormal{-}small}$ and $\mathrm{Straggler\textnormal{-}relaunch}$ require the task service times in order to decide whether to schedule a job with redundancy or to set its relaunch time.
However, service times might be unknown or only known with some degree of uncertainty in practice.
Service times in general are desired to be known or estimated well enough for effective management of cluster resources and to achieve good performance from the deployed job scheduling policies, and statistical techniques have been applied to predict service times of the arriving jobs in cloud or HPC workloads \cite{WorkloadModelingForCloud:GanapathiCF10, PredictionOfJobDurationsInHPC:GalleguillosSK17}.
Even then the predictions might be inaccurate to a degree that would significantly disrupt the performance expected from the scheduling policy.
There has been a recent interest in understanding the effects of uncertainty in the service times on the performance of service scheduling policies (such as ``Shortest remaining processing time'' first) in $M/G/1$ queue \cite{RobustSchedulingUnderAdversarialNoise:ScullyH18, SchedulingWithPredictions:Mitzenmacher19}. 
Scheduling policies, such as the two studied in this paper for job execution with redundancy and/or straggler relaunch, should also be studied along this line in order to understand the impact of mispredictions or inaccuracy in predicting task service times on the scheduling performance.

\vspace{0.5em}
\noindent
\textbf{$\mathbf{M/G/c}$ approximation.}
In Approximation~\ref{approx_MGc}, we proposed that Master-Worker compute system under a work conserving policy (such as $\mathrm{Redundant\textnormal{-}small}$ and $\mathrm{Straggler\textnormal{-}relaunch}$) can be approximated as an $M/G/c$ queue with a properly adjusted service time distribution and number of servers $c$, both of which we found by using the first and/or second moments for the latency and cost of an arbitrary job execution.
Then, we used the well known approximation available for the average response time in $M/G/c$ queue in order to approximate the average response time $\E[T]$ in the Master-Worker system under $\mathrm{Redundant\textnormal{-}small}$ (in Sec.~\ref{sec:sec_redsmall}) or $\mathrm{Straggler\textnormal{-}relaunch}$ (in Sec.~\ref{sec:sec_relaunch}) policy.
We found that the approximation \eqref{eq:eq_ET_approx} for $\E[T]$ allows optimizing the performance of $\mathrm{Redundant\textnormal{-}small}$ and $\mathrm{Straggler\textnormal{-}relaunch}$ policies fairly accurately.
However, \eqref{eq:eq_ET_approx} uses the second moment of latency, which will become $\infty$ once the tail index of task service time distribution $\beta \leq 2$ because the second moment of $\mathrm{Pareto}$ distribution is $\infty$ when its tail index is $\leq 2$.
Thus, in order to use \eqref{eq:eq_ET_approx} for workloads with a very heavy tail, we need to model the task service time distribution $B$ with a Truncated-Pareto distribution, which has all moments finite for any tail index value and shown to fit real job sizes \cite{UnfairnessSRPT:BansalH01}.

Besides the approximation that we use for $\E[T]$ in $M/G/c$ queue, there are numerous others available in the literature \cite{ApproximationsForMGcQ:Hokstad78, ApproximationsForMGcQ:BoxmaCH79, MGcUnderHeavyTailedWorkload:PsounisMP05}.
Our choice of the approximation is mainly motivated by its simplicity, and other approximations might give better overall accuracy in estimating and tuning the performance of Master-Worker compute system.
Approximation that we used for $\E[T]$ in $M/G/c$ queue is expected to decline in accuracy as the variability in task service times increases \cite{ApproximationsForGGcQ:Whitt93, OnTheInapproximabilityOfMGc:GuptaHD10}.
Furthermore, the authors in \cite{OnTheInapproximabilityOfMGc:GuptaHD10} rigorously show that any approximation for the $\E[T]$ in $M/G/c$ queue based on only the first two moments of the job service time will be inaccurate for some job service time distribution.
We use the approximate results for the performance of $M/G/c$ queue in estimating the performance of Master-Worker system with multi-task job arrivals and job scheduling with redundancy or straggler relaunch policies.
Thus, as the approximate expression for $M/G/c$ declines in accuracy, we expect \eqref{eq:eq_ET_approx} to also get worse in estimating the $\E[T]$ in the Master-Worker system.
We observe that this is indeed the case in the simulations (which are omitted here due to space constraint).
However, despite the discrepancy between the simulated and estimated values, both the simulations and the approximation yield very similar trajectories of increase or decrease in $\E[T]$ as we increase $d$ (the threshold on job demand to schedule with redundancy in $\mathrm{Redundant\textnormal{-}small}$ policy) or $w$ (multiplicative factor that determines the relaunch time of jobs in $\mathrm{Straggler\textnormal{-}relaunch}$ policy).
Thus, $M/G/c$ approximation is still able to guide us to find fairly accurate estimates of the optimal value of $d$ or $w$ under varying levels of offered load, even when the tail of task service times is very heavy tailed (i.e., $\beta < 2$).

There is a large literature on estimating the performance of $M/G/c$ queue under different workload and scheduling models. We demonstrated that approximating a Master-Worker system with multi-task job arrivals as an $M/G/c$ queue is promising to yield important insight into the system performance under practical scheduling policies such as $\mathrm{Redundant\textnormal{-}small}$ or $\mathrm{Straggler\textnormal{-}relaunch}$, and for optimizing the performance of such policies based on the straggling and workload characteristics.
Further investigation of this approach via more rigorous arguments and other techniques available in the literature would be fruitful to derive scheduling policies with redundancy or straggler relaunch that perform well in practice.

\section{Conclusion}
This paper is one of the first to address the problem of scheduling in Master-Worker compute systems that use redundancy to mitigate stragglers.
We found optimal scheduling policies in two stages.
We firstly used RL techniques to learn the principles for effective scheduling of redundancy.
Then building on these principles, we proposed a simple policy with mathematical modeling and presented an approximate analysis of its performance.
We observed that our policy performs as good as the more complex policies that could possibly be learned by parameter-optimized Deep-RL alone.
We extended our approximate analysis when the stragglers are mitigated by relaunching them rather than employing redundant tasks.
Our approximate analysis allows tuning the parameters of both the policy with redundancy and the policy with relaunching stragglers, and we found that optimized policy with redundancy significantly outperforms straggler relaunch when the offered load on the system is low or moderate, and performs worse when the offered load is very high ($\gtrsim 0.85$).

\newpage
\balance
\bibliographystyle{unsrt}
\bibliography{references}

\end{document}